\documentclass[onecolumn]{mn2e}
\usepackage{epsfig}
\usepackage{bm}
\usepackage{colordvi}
\usepackage{color}
\def\beq#1{\begin{equation}\label{#1}}
\def\eeq{\end{equation}}
\def\beqa#1{\begin{eqnarray}\label{#1}}
\def\eeqa{\end{eqnarray}}

\def\h0{H$_{\rm 0}$~}

\title[The Dark Energy Equation of State ]{High Redshift Investigation On The Dark Energy Equation of State } \author[E.Piedipalumbo$^{1,2}$, E. Della Moglie$^{3}$, M. De Laurentis$^{1,2}$, P. Scudellaro$^{1,2}$]{E.Piedipalumbo$^{1,2}$, E. Della Moglie$^{3}$, M. De Laurentis$^{1,2}$,  P. Scudellaro$^{1,2}$\\
$^1$ Dipartimento di Fisica, Universit\`{a} di Napoli
Federico II, Compl.
Univ. Monte S. Angelo, 80126 Napoli, Italy  \\
$^2$ I.N.F.N., Sez. di Napoli, Complesso Universitario di Monte
Sant' Angelo, Edificio G, via Cinthia, 80126 Napoli, Italy\\
$^3$ Dipartimento di Ingegneria della Produzione,Termoenergetica e Modelli Matematici, Universit\`{a} di Genova,\\\,\, P.le J.F. Kennedy, I 16129  Genova}
\date{Accepted xxx, Received yyy, in original form zzz}
\begin{document}
\maketitle
\begin{abstract}
The understanding of the accelerated expansion of the Universe poses one of
the most fundamental questions in physics and cosmology today. Whether or not the acceleration is driven by
some form of dark energy, and in the absence of a well-based theory to interpret the observations, many models have been
proposed to solve this problem, both in the context of General Relativity and alternative theories of gravity. Actually, a further possibility to investigate the nature of dark energy lies in measuring the dark energy equation of state (EOS) , $w$, and its time (or redshift) dependence at high accuracy. However, since $w(z)$ is not directly accessible to measurement, reconstruction methods are needed to extract it reliably from observations. Here we investigate different models of dark energy, described through several parametrizations of the EOS. Our high-redshift analysis is based on the Union2 Type Ia Supernovae (SNIa) data set, the Hubble diagram constructed from some Gamma Ray Bursts (GRBs) luminosity distance indicators, and Gaussian priors on the distance from the Baryon Acoustic Oscillations (BAO), and the Hubble constant $h$ (these priors have been included in order to help break the degeneracies among model parameters). To perform our statistical analysis and to explore the probability distributions of the EOS parameters we use the Markov Chain Monte Carlo Method (MCMC). It turns out that the dark energy equation  of state is evolving for all the parametrizations that we considered.
We finally compare our results with the ones obtained by previous cosmographic analysis performed on the same astronomical datasets, showing that the latter ones are sufficient to test and compare the new parametrizations.
\end{abstract}

\section{Introduction}
It is well known that by the end of the nineties, from observations of supernovae at high redshift, the Universe is expanding. The observations of scale temperature anisotropies of the Cosmic Microwave
Background radiation (CMB)  have confirmed this result independently \cite{Riess07,SNLS,Union,WMAP3,PlanckXXVI}.
It is common practice to assume that the observed accelerated expansion is caused by dark energy  with unusual properties. The pressure of
dark energy $p_{de}$ is negative and it is related to the positive
energy density of dark energy $\epsilon_{de}$ by
$p_{de}=w\epsilon_{de}$ where the proportionality coefficient $w<0$.
Even today, the nature of dark energy is unknown, we only know that was estimated to be about $75\%$ of matter-energy
in the Universe and  its properties are characterized by the
EOS parameter, $w$. Extracting the information
on EOS of dark energy from observational data is then  a fundamental problem. Get informations from the observed data on the EOS of dark energy is at same time an issue of crucial importance and a challenging task.
For probing the dynamical evolution of dark energy, under such circumstances, one can parameterize $w$ empirically, usually assuming that this quantity evolves smoothly with redshift, so that it can be approximated by a fitting analytical expression, using two or more free parameters.
Among all the parametrization forms of EOS,  we will consider
the Chevallier-Polarski-Linder (CPL) model~\cite{cpl1,cpl2}, which is widely used, since it presents  a well behaved and bounded
behavior for high redshifts, and a manageable two-dimensional parameter space.
However, we will also introduce new parametrizations, that have been recently introduced by \cite{novel} and \cite{oscisalz} to avoid the divergency problem inherent to theCPL parametrization, which turned out to be able to satisfy many theoretical scenarios.
For constraining the parameters, which appear in the EOS, we use  a large collection of cosmological datasets: the Union2 Type Ia SNIa data set, the Hubble diagram constructed from some GRBs luminosity distance indicators, and  in order to help break the degeneracies among model parameters, Gaussian priors on the distance from the BAO, and the Hubble constant $h$.  Actually, observations of the SNIa are consistent with the assumption that the observed accelerated expansion is due to the non zero cosmological constant.
However, so far the SNIa have been observed only at redshifts $z<2$, while in order to test if $w$ is changing with redshift it is necessary to use more distant objects. New possibilities opened up when the GRBs have been discovered at  higher redshifts. The discovery of GRBs at higher redshift has opened up new avenues for cosmology, although they remain enigmatic objects.
  First of all, the mechanism that is responsible for releasing the incredible amounts of energy that a typical GRB emits is not yet known (see for instance Meszaros 2006 for a recent review).  It is also not yet definitely known if the energy is emitted isotropically or is beamed. Despite these difficulties, GRBs are promising objects that can be used to study the expansion rate of the Universe at high redshifts \cite{Bradley03,S03,Dai04,Bl03,Firmani05,S07,Li08,Amati08,Ts09}.

 Actually, even if the huge dispersion (about four orders of magnitude) of the isotropic GRB energy makes them everything but standard candles, it has been recently empirically established that some of the directly observed parameters of GRBs are correlated with their important intrinsic parameters, like the luminosity or the total radiated energy, allowing to derive some correlations, which have been tested and used to calibrate such relations, and to derive
their luminosity or radiated energy from one or more observables, in order to construct a GRBs Hubble diagram. It has been shown that such a procedure can be implemented without specifying the cosmological model; see, for instance, \cite{MEC10,ME} and references therein.
In our analysis we use a GRB HD data set consisting of $109$ high redshift GRBs, which has been constructed from the Amati $E_{\rm p,i}$ -- $E_{\rm iso}$ correlation (here $E_{\rm p,i}$ is the peak photon energy of the intrinsic spectrum and $E_{\rm iso}$ the isotropic equivalent radiated energy), applying a local regression technique to estimate, in a model independent way,
the distance modulus from the recently updated Union SNIa data set.
It turns out that these data sets are sufficient for our aim of testing and comparing the new parametrizations.

The scheme of the paper is as follows. In Section 2 we describe the basic elements of the parametrizations of the considered EOS, while in Section 3 we introduce the observational data sets that are used in our analysis. In Section 4 we describe some details of our statistical analysis from three sets of data. In a general discussion of our results and conclusions in Section 5, we finally present some constrains on dark energy models that can be derived from our analysis.

\section{ Dark Energy Parametrizations }

The discovery from the SNIa that the expansion rate of the Universe is apparently accelerated is one of the most significant events in the modern cosmology. Although seemingly consistent with our current concordance model in which the source of the cosmic acceleration takes the form of the Einstein's cosmological constant, the precision of current data is not sufficient to rule out the possibility of an evolving component. If then the $\Lambda$CDM model is not correct, we are perhaps looking for some dynamical field with a repulsive gravitational force. Moreover this could instead be indicating that the Copernican principle is wrong, and that radial inhomogeneity is responsible for the {\textit accelerated expansion}. Within the Friedmann-Lemaitre-Robertson-Walker (FLRW) paradigm, all possibilities can be characterized, as far as the background dynamics are concerned, by the dark energy EOS $w(z)$.
Unfortunately, from a theoretical perspective  $w(z)$ could really be pretty much anything. A priority in cosmology today lies in searching for evidence for $w(z)\neq-1$.

The observational challenge to solve such ambiguity lies in finding a general way to treat $w(z)$.
This is usually done in terms of a simple parameterization of $w(z)$; but any such functional forms for $w(z)$ are problematic because they have no basis in a grounded  theory and to be flexible could require a large set of parameters. However at the present the signal to noise ratio in the observational data  is not enough to provide constraints in more than few parameters (two or three at most). To  reduce the huge arbitrariness, the space of allowed $w(z)$ models is often reduced to $w\geq-1$; however when  $w$ is an effective EOS, parameterizing a modified gravity theory, as for instance a scalar tensor or a $f(R)$-model \cite{PRnostro}, then this constraint  might be too restricted.
An alternative procedure is to reconstruct $w(z)$ directly from the observables without any dependence on a parameterization of $w(z)$ or understanding of dark energy, as done, for example in~\cite{SF}.
Some direct reconstruction methods rely, for instance,  on estimating the first and second derivatives of luminosity-distance data.
Actually, defining  $D(z)=(H_0/c)(1+z)^{-1}d_L(z)$, it turns out that

\begin{equation}
w(z)=\frac{2(1+z)(1+\Omega_kD^2)D''-\left[(1+z)^2\Omega_k D'^2+2(1+z)\Omega_k D D'-3(1+\Omega_kD^2)\right]D'}{3\left\{(1+z)^2\left[\Omega_k+(1+z)\Omega_m\right]D'^2-(1+\Omega_kD^2)\right\}D'}.
\label{w}
\end{equation}

Thus, given a  a parameterized ansatz for $D(z)$, it is possible to  reconstruct the dark energy EOS from Eq~(\ref{w})\,\,. See for instance~\cite{SF}, and references therein, for a review,
and~\cite{phenix,DalyDjorgovski2,Clarkson:2010bm,Lazkoz:2012eh,said} for an overview about critical topics and alternative model independent approaches connected to the dark energy reconstruction techniques.
New and interesting  prospectives to extract information of the dark energy modeling based on a recent approach, the so called Genetic Algorithms are illustrated in  \cite{nesdark,disera,disera1}.
 Here we are investigating  if, by analyzing a large collection of  cosmological data, any indications of a deviation from the $w(z)\neq-1$ come to light, as  we detected in a previous cosmographic analysis, where the value of the  deceleration parameter clearly confirmed the present acceleration phase, and the estimation of the jerk  reflected the possibility of a deviation from the $\Lambda$CDM cosmological model. To accomplish this task we focus on  a direct and full reconstruction of  the dark energy EOS through several parameterizations, widely used in literature.

\subsection{Basic equations}

 Within the FLRW paradigm, dark energy appears in the Friedmann equations of
cosmological dynamics through its effective energy density
and pressure:

\begin{equation}
\frac{\ddot{a}}{a} = - \frac{4 \pi G}{3} \ \left(\rho_M + \rho_X + 3 p_X\right) \ ,
\label{eq: fried1}
\end{equation}

\begin{equation}
H^2 = \frac{8 \pi G}{3} (\rho_M + \rho_X) \,.
\label{eq: fried2}
\end{equation}
Here $a$ is the scale factor, $H = \dot{a}/a$ the Hubble parameter, the dot denotes the derivative with respect to cosmic time, and we have assumed a spatially flat Universe in agreement with what is inferred from CMBR anisotropy spectrum \cite{PlanckXXVI}.
The continuity equation for any cosmological fluid is \,:

\begin{equation}
\frac{\dot{\rho_i}}{\rho_i} = - 3 H \left(1 + \frac{p_i}{\rho_i}\right) = - 3 H  \left[1 + w(t)\right]\,,
\label{eq: continuity}
\end{equation}
where the energy density is $\rho_i$, the pressure $p_i$, and the
EOS of each component is defined by
$\displaystyle{w=\frac{p_i}{\rho_i}}$.
Ordinary nonrelativistic matter has $w=0$, and the
cosmological constant has $w=-1$.
If we explicitly allow the possibility that the dark energy evolves, the importance its equation of state is significant and determines the expression of the Hubble function $H(z)$, and any derivation of it needed to obtain the observable quantities. Actually it turns out that:
\begin{eqnarray}\label{heos}
  H(z,{\bm\theta}) &=& H_0 \sqrt{(1-\Omega_m) g(z, {\bm \theta})+\Omega_m (z+1)^3}\,,\\
\end{eqnarray}
 where $g(z)=\frac{\rho_{de}(z)}{\rho_{de}(0)}=\exp^{3 \int_0^z \frac{w(x,{\bm \theta})+1}{x+1} \, dx}$,  $w(z,{\bm\theta})$ is any dynamical form of the dark energy EOS, and ${\bm \theta}=(\theta_1, \theta_1..,\theta_n)$ are the dark energy EOS  parameters.
Moreover
\begin{eqnarray}
  d_L(z,{\bm\theta}) &=& \frac{c}{H_0} (1+z)\int_0^z \frac{1}{  \sqrt{(1-\Omega_m) g(y,{\bm \theta})+\Omega_m (y+1)^3}}dy,\label{lumd} \\
  d_A(z,{\bm \theta}) &=&  \frac{c}{H_0}\frac{1}{1+z} \int_0^z \frac{1}{  \sqrt{(1-\Omega_m) g(y,{\bm\theta})+\Omega_m (y+1)^3}}dy,\label{angd} \\
  d_V(z,{\bm \theta}) &=& \left[\left(1+z\right) d_A(z,{\bm \theta})^2\frac{c z}{H(z,{\bm\theta})}\right]^{\frac{1}{3}},\label{volumed}
\end{eqnarray}
where  $ d_L(z,\mathbf{\theta})$ is the luminosity distance, $ d_A(z,{\bm\theta})$ the angular diameter distance and $ d_V(z,\mathbf{\theta})$ the volume distance defined by \cite{Eis05}\,:. All of them are needed to perform our statistical analysis.   In this work we consider three different parametrizations:
\begin{itemize}
 \item the so-called Chevalier-Polarski Linder (CPL) model \cite{cpl1,cpl2}.This parametrization
assume a dark energy EOS given by
\begin{equation}
w(z) =w_0 + w_{1} z (1 + z)^{-1} \,,
\label{cpleos}
\end{equation}
where $w_0$ and $w_a$ are real numbers that represent the EOS present value and its
overall time evolution, respectively \cite{cpl1,cpl2}.
 It is important to remember that for high redshift we have the following behavior
\begin{equation}
\lim_{z \to \infty}w^{CPL}(z)=w_0+w_a =: w_i^{CPL}
\end{equation}
that allows us to describe a wide variety of scalar field  dark energy models. Then, this parameterization appears to be a
good compromise to construct a model independent analysis.

\item  a novel parametrization recently introduced in \cite{novel} to avoid the future divergency
problem of the CPL parametrization, and to probe the
dynamics of dark energy not only in the past evolution but also
in the future evolution,
\begin{equation}\label{noveleos}
w(z)=w_0+w_1 \left[\frac{\sin (z+1)}{z+1}-\sin (1)\right]\,,
\end{equation}

\item an oscillating dark energy EOS recently discussed in \cite{oscisalz}
\begin{equation}
 w(z)=\frac{w_1 \left[1-\cos (\delta  \log (z+1))\right]}{\log (z+1)}+w_0\,.\label{osceos}
\end{equation}
These  oscillating models have been proposed to solve the so called {\it coincidence problem} very easily, due to the sequence of different periods of acceleration, and are available in several theoretical scenarios.
\end{itemize}

It is worth noting that, it is possible to build up the link between the dark energy EOS and the cosmographic parametrization (based on the series expansion (in redshift) of the Hubble function $H(z)$),in order to finally cross-correlate the results obtained from such independent approaches. Actually it turns out that, at fourth order (in $z$):
\begin{eqnarray}
\frac{dg}{dz}\Big|_{z=0} ({\bm \theta}) &=& \frac{2 - 3 \Omega_m + 2 q_0 }{1-\Omega_m}\,, \\
\frac{d^2g}{dz^2}\Big|_{z=0}({\bm\theta}) &=& \frac{-2 \left(-1 - j_0 + 3 \Omega_m - 2 q_0\right)}{1-\Omega_m}  \,, \\
\frac{d^3g}{dz^3}\Big|_{z=0}({\bm \theta}) &=&\frac{ -2 \left(3 \Omega_m + j_0 q_0 + s_0\right)}{1-\Omega_m}  \,, \\
\frac{d^4G}{dz^4}\Big|_{z=0}({\bm\theta}) &=& \frac{1}{1-\Omega_m}\times 2 \bigl(-12 j_0 + 3 j_0^2 + 12 j_0 \Omega_m - 4 j_0^2 \Omega_m + l_0 \Omega_m - 28 j_0 q_0 +
   32 j_0 \Omega_m q_0 + 12 q_0^2 - 22 j_0 q_0^2 + \\
  &&-\, 12 \Omega_m q_0^2 + 25 j_0 \Omega_m q_0^2 + 24 q_0^3 - 24 \Omega_m q_0^3 + 15 q_0^4 -
  15 \Omega_m q_0^4 - 4 s_0 + 8 \Omega_m s_0 - 4 q_0 s_0 + \\
  &&+\, 7 \Omega_m q_0 s_0 + (1 - \Omega_m) (-4 j_0^2 + l_0 - 12 q_0^2 - 24 q_0^3 - 15 q_0^4 +
       j_0 (12 + 32 q_0 + 25 q_0^2) + 8 s_0 + 7 q_0 s_0)\bigr) \,,
        \label{cosmoeos}
\end{eqnarray}
where the function $g(z,\mathbf{\theta})$ is defined above, and $q_0,j_0, l_0$ are the present values of the following cosmographic functions

\begin{eqnarray}\label{eq:cosmopar}
q(t) &\equiv& - \frac{1}{a}\frac{d^{2}a}{dt^{2}}\frac{1}{H^{2}}\,
\\
j(t) &\equiv& + \frac{1}{a}\frac{d^{3}a}{dt^{3}}\frac{1}{H^{3}}\,
,
\\
s(t) &\equiv& + \frac{1}{a}\frac{d^{4}a}{dt^{4}}\frac{1}{H^{4}}\,
,
\\
l(t) &\equiv& + \frac{1}{a}\frac{d^{5}a}{dt^{5}}\frac{1}{H^{5}}\,.
\end{eqnarray}
It is worth noting that the deceleration parameter $q(z)$ can be related to the EOS through the Hubble parameter $H(z)$
\begin{eqnarray}
&&1+q(z) = \epsilon(z) = - {\dot H\over H^2} = (1+z)\frac{H'(z)}{H(z)} =\frac{d \ln H(z)}{d \ln(1+z)}\\ \label{q(z)}
&&=-\frac{H_0 (z+1) \left(3 (z+1)^2 \Omega _m-\left(\Omega _m-1\right) \frac{d g}{dz}(z,{\bm\theta})\right)}{2 \sqrt{(z+1)^3 \Omega
   _m-\left(\Omega _m-1\right) g(z,{\bm\theta})}}\,,\nonumber
\end{eqnarray}

\section{Observational data sets}
\label{s3}

In our  approach we use a great collection of presently available observational data sets on SNIa and GRB Hubble Diagrams, and we set Gaussian priors on the distance from the BAO, and the Hubble constant $h$.
\label{sec:SNdata}
 Over the last decade the confidence in SNIa as standard candles has been steadily
growing.  Actually, the SNIa observations gave the first strong indication of an accelerating
expansion of the Universe, which can be explained by assuming the existence of some kind of dark energy or
nonzero cosmological constant. Since $1995$ two teams of astronomers - the High-Z
Supernova Search Team and the Supernova Cosmology Project - have been discovering SNIa at high
redshifts.
Here we consider the recently updated Supernovae Cosmology Project \textit{Union 2.1} compilation
\cite{Union21}, which is an update of the original \textit{Union} compilation, consisting of
 $580$ SNIa, spanning the redshift range ($0.015 \le z \le 1.4$). We actually
compare the \textit{theoretically\thinspace\ predicted} distance modulus $\mu(z)$
with the \textit{observed} one, through a Bayesian approach, based on the definition
of the distance modulus,
\begin{equation}
\mu(z_{j}) = 5 \log_{10} ( D_{L}(z_{j}, \{\theta_{i}\}) )+\mu_0\,,
\end{equation}
where $D_{L}(z_{j}, \{\theta_{i}\})$ is the Hubble free luminosity
distance, expressed as a series depending on the EOS parameters, $\theta_{i}=(w_0,...w_{i}..)$, and  $\mu_{0}$ encodes the Hubble
constant and the absolute magnitude $M$.

\subsection{GRBs Hubble diagram}

 GRBs are visible up to high $ z $, thanks to the enormous energy that they release, and thus may be good candidates  for our high-redshift cosmological investigation. Sadly, GRBs may be everything but standard candles since their peak luminosity spans a wide range, even if  there have been many   efforts to make them standardizable candles using
some empirical correlations among distance dependent quantities and rest frame observables  \cite{Amati08}.
 These empirical relations allow one to deduce the GRB rest frame luminosity or energy from an observer
frame measured quantity so that the distance modulus can be obtained with an error which depends essentially on the intrinsic scatter of the adopted correlation.

 Combining the estimates from different correlations, \cite{S07}
first derived the GRBs HD for $69$ objects, which has been further enlarged using updated samples, different calibration methods and also different correlation relations, see for instance
\cite{MEC10,ME}, showing the interest in the cosmological applications of GRBs. In this paper we perform our cosmographic analysis using two GRBs HD data set, build up by calibrating the Amati $E_{\rm p,i}$ -- $E_{\rm iso}$ .

\subsubsection{The \textit{calibrated Amati} Gamma Ray Bursts Hubble diagram}

Recently it has been empirically established that some of the directly observed parameters of GRBs
are connected with the isotropic absolute luminosity $L_{iso}$, or the isotropic bolometric energy $E_{\rm iso}$ of a GRB.
These quantities appear to correlate with the GRB isotropic luminosity, its total collimation-corrected or its
isotropic energy. The isotropic luminosity and energy can  not be measured directly but rather it can be obtained through the knowledge
of either the bolometric peak flux, denoted by $ P_{bolo} $, or the bolometric fluence, denoted by $ S_{bolo} $.
Actually
\begin{equation}
 L_{\rm iso} = 4 \pi d^2_{L}(z) P_{\rm bolo} \,,
 \label{ldl}
\end{equation}
and
\begin{equation}
E_{\rm iso}=4\pi d^2_{L}(z)S_{\rm bolo}(1+z)^{-1}\,.
\end{equation}

\begin{figure}
\includegraphics[width=8 cm, height=6 cm]{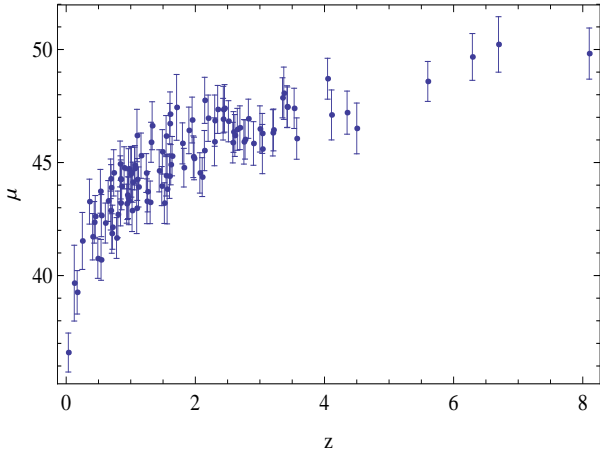}
\caption{Distance modulus $\mu(z)$ for the \textit{calibrated}  GRBs Hubble diagram made up by fitting the Amati
correlation.} \label{figrbs}
\end{figure}

Therefore, $L_{iso}$, and $ E_{\rm iso}$ depend  on the GRB observables , $P_{\rm bolo}$ and
$S_{\rm bolo}$, but also on the cosmological parameters. Therefore, at a first glance it seems impossible to calibrate such GRBs
empirical laws, without assuming any a priori cosmological model. This is the so called circularity problem, which has to be overcome, in order to use GRBs as tools for cosmology.
In  \cite{ME} we have applied a local regression technique to estimate, in a model independent way, the distance modulus from the
Union SNIa sample, containing $580$ SNIa spanning the redshift range of $0.015 \le z \le 1.4$.  In particular, by using such
a technique, we have fitted the so-called \textit{Amati relation} and constructed an updated GRBs
Hubble diagram, which we call the \textit{calibrated} GRBs HD, consisting of a sample of $109$ objects, shown in Fig. \ref{figrbs}.
While both SNIa and GRBs are based on the concept of standard candles, even if for the GRBs such a concept is {\it generalized} toward an  unorthodox meaning, an alternative way to probe the background evolution of the Universe relies on the use of standard rulers.
Nowadays the BAOs which are related to the imprint of the primordial acoustic waves on the galaxy power spectrum are widely used as such rulers.
In order to use BAOs as constraints, we follow \cite{P10} by first defining\,:
\begin{equation}
d_z = \frac{r_s(z_d)}{d_V(z)}\,,
\label{eq: defdz}
\end{equation}
with $z_d$ the drag redshift computed using the approximated formula in \cite{EH98}, $r_s(z)$ the comoving sound horizon given by\,:
\begin{equation}
r_s(z) = \frac{c}{\sqrt{3}} \int_{0}^{(1 + z)^{-1}}{\frac{da}{a^2 H(a) \sqrt{1 + (3/4) \Omega_b/\Omega_{\gamma}}}} \ ,
\label{defsoundhor}
\end{equation}
and $d_V(z)$ the volume distance defined in (\ref{volumed})\,.

\section{Statistical Analysis}

In this section we show our statistical analysis and present our main results on the constraints for the
  EOS parameters from the current observational data sets described above. In order to constrain the parameters, describing each of the selected dark energy EOS, we perform a preliminary and \textit{standard fitting} procedure to maximize the likelihood function ${\cal{L}}({\bf p}) \propto \exp{[-\chi^2({\bf p})/2]}$, where ${\bf p}$ is the set of cosmographic parameters and the expression for $\chi^2({\bf p})$ depends on the data set used.  As a first test we consider only the SNIa data, thus we define\,:
\begin{eqnarray}
\chi^2({\bf p}) & = & \sum_{i = 1}^{{\cal{N}}_{SNIa}}{\left [ \frac{\mu_{obs}(z_i) - \mu_{th}(z_i, {\bf p})}{\sigma_i} \right ]^2} \nonumber \\
& + &  \left ( \frac{h - 0.742}{0.036} \right )^2
+ \left ( \frac{\omega_m - 0.1356}{0.0034} \right )^2 \ .
\label{defchiSNIa}
\end{eqnarray}
Here, $\mu_{obs}$ and $\mu_{th}$ are the observed and theoretically predicted values of the distance modulus, while the sum is over all the SNIa  in the sample. The last two terms are Gaussian priors on
$h$ and $\omega_M = \Omega_M h^2$ and are included in order to help break the degeneracies among the model parameters.
We have resorted to the results of the SHOES collaboration \cite{shoes} and the WMAP7 data \cite{WMAP7}, respectively, to set the numbers used in Eqs. (\ref{defchiSNIa}).
When we are using GRBs only, we define\,:
\begin{eqnarray}
\chi^2({\bf p}) & = & \sum_{i = 1}^{{\cal{N}}_{GRBHD}}{\left [ \frac{\mu_{obs}(z_i) - \mu_{th}(z_i, {\bf p})}{\sigma_i} \right ]^2} \nonumber \\
& + &  \left ( \frac{h - 0.742}{0.036} \right )^2
+ \left ( \frac{\omega_m - 0.1356}{0.0034} \right )^2 \ .
\label{chigrb}
\end{eqnarray}

As a next step, we combine the SNIa and GRBs HDs with other data redefining ${\cal{L}}({\bf p})$ as\,:
\begin{eqnarray}
{\cal{L}}({\bf p}) & \propto & \frac{\exp{(-\chi^2_{SNIa/GRB}/2)}}{(2 \pi)^{\frac{{\cal{N}}_{SNIa/GRB}}{2}} |{\bf C}_{SNIa/GRB}|^{1/2}} \nonumber \\
~ & \times  & \frac{1}{\sqrt{2 \pi \sigma_h^2}} \exp{\left [ - \frac{1}{2} \left ( \frac{h - h_{obs}}{\sigma_h} \right )^2
\right ]} \nonumber \\
~ & \times & \frac{\exp{(-\chi^2_{BAO}/2})}{(2 \pi)^{{\cal{N}}_{BAO}/2} |{\bf C}_{BAO}|^{1/2}} \nonumber \\
~ & \times & \frac{1}{\sqrt{2 \pi \sigma_{{\cal{R}}}^2}} \exp{\left [ - \frac{1}{2} \left ( \frac{{\cal{R}} - {\cal{R}}_{obs}}{\sigma_{{\cal{R}}}} \right )^2 \right ]} \nonumber \\
~ & \times & \frac{\exp{(-\chi^2_{H}/2})}{(2 \pi)^{{\cal{N}}_{H}/2} |{\bf C}_{H}|^{1/2}} \  .
\label{defchiall}
\end{eqnarray}

\begin{figure}
\includegraphics[width=6 cm, height=6 cm]{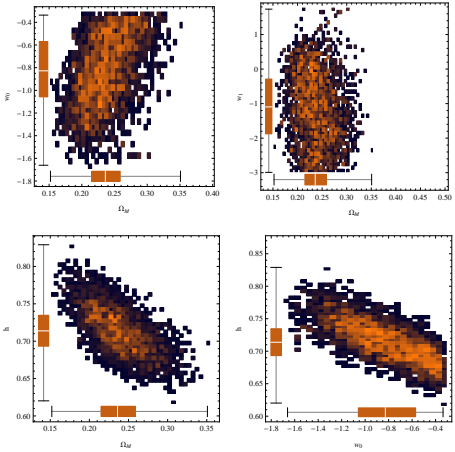}
\caption{The joint probability for different couples of parameters which characterize the CPL EoS, as provided by  our analysis. On the axes are plotted the box-and-whisker diagrams relatively to the different parameters: the bottom and top of the diagrams are  the 25th and 75th percentile (the lower and upper quartiles, respectively), and the band near the middle of the box is  the 50th percentile (the median).} \label{likecpl}
\end{figure}

The first two terms are the same as above with ${\bf C}_{SNIa/GRB}$ the SNIa/GRBs diagonal covariance matrix and $(h_{obs}, \sigma_h) = (0.742, 0.036)$. The third term takes into account the constraints on $d_z = r_s(z_d)/D_V(z)$ with $r_s(z_d)$ the comoving sound horizon at the drag redshift $z_d$ (which we fix to be $r_s(z_d) = 152.6 \ {\rm Mpc}$ from WMAP7) and the volume distance is defined as in Eq. (\ref{volumed}).
The values of $d_z$ at $z = 0.20$ and $z = 0.35$ have been estimated by Percival et al. (2010) using the SDSS DR7 galaxy sample so that we define $\chi^2_{BAO} = {\bf D}^T {\bf C}_{BAO}^{-1} {\bf C}$ with ${\bf D}^T = (d_{0.2}^{obs} - d_{0.2}^{th}, d_{0.35}^{obs} - d_{0.35}^{th})$ and ${\bf C}_{BAO}$ is the BAO covariance matrix. The next term refers to the shift parameter \cite{B97,EB99}\,:

\begin{equation}
{\cal{R}} = H_{0} \sqrt{\Omega_M} \int_{0}^{z_{\star}}{\frac{dz'}{H(z')}}\,,
\label{eq: defshiftpar}
\end{equation}
with $z_\star = 1090.10$ the redshift of the last scattering surface. We follow again the WMAP7 data setting $({\cal{R}}_{obs}, \sigma_{{\cal{R}}}) = (1.725, 0.019)$. While all these quantities (except for the Gaussian prior on $h$) mainly involve the integrated $E(z)$, the last term refers to the actual
measurements of $H(z)$ from the differential age of passively evolving elliptical galaxies.
We then use the data collected by Stern et al. (2010) giving the values of the Hubble parameter for ${\cal{N}}_H = 11$ different points over the redshift range $0.10 \le z \le 1.75$ with a diagonal covariance matrix.
We finally perform our statistical analysis, considering a whole data set containing both the SNIa Union data set and the \textit{calibrated} GBRs HD, and slightly modifying the likelihood ${\cal{L}}({\bf p})$.
Actually,  in order to efficiently sample the ${\cal{N}} $ dimensional parameter space, we use the Markov Chain Monte Carlo (MCMC) method running five parallel chains and using the Gelmann - Rubin convergence test. It is worth noting that  the Gelman-Rubin diagnostic uses parallel chains with dispersed initial values to test whether they all converge to the same target distribution. Failure could indicate the presence of a multi-mode posterior distribution (different chains converge to different local modes) or the need to run a longer chain.
As a test instrument it uses the reduction factor \textit{R}, which is the square root of the ratio of the between-chain variance and the within-chain variance. A large \textit{R} indicates that the between-chain variance is substantially greater than the within-chain variance, so that longer simulation is needed.
We want that  \textit{R}  converges to 1 for each parameters. We set $R - 1$ of order $0.05$, which is more restrictive than the often used and recommended value $R - 1 < 0.1$ for standard cosmological investigations.
We test the convergence of the chains by the Gelman and Rubin criterion,
Moreover in order to reduce the uncertainties on EOS parameters, since methods like the MCMC are based on an algorithm that moves randomly in the parameter space, we \textit{a priori} impose some basic consistency constraints on the positiveness of $H^{2}(z)$ and $d_{L}(z)$.
We first run our chains to compute the likelihood in Eqs. (\ref{defchiSNIa}) and/or (\ref{chigrb}), using as starting points the best fit values obtained  in our \textit{pre-statistical analysis}, in order
to \textit{select} the starting points. Therefore we perform the same Monte Carlo Markov Chain calculation to evaluate the likelihood in Eq. (\ref{defchiall}), combining the
SNIa HD, the BAO and $H(z)$ data with the GRBs HD respectively, as described above.
We throw away  first $30\%$  of the points iterations at the beginning of any MCMC run, and we thin the many times -runned chains.
We finally   extract the constraints on EOS parameters, coadding the thinned chains.  The histograms of the parameters from the merged chain after burn in cut and thinning are then used to infer median values and confidence ranges. Actually, the  confidence levels are estimated
from the final sample (the merged chain): the $15.87$-th and $84.13$-th quantiles define the $68 \%$ confidence interval; the $2.28$-th and $97.72$-th quantiles define the $95\%$ confidence interval;  and the $0.13$-th and $99.87$-th quantiles
define the $99\%$ confidence interval. In Table \ref{tab:fulltab} we present the results of our analysis. It turns out that for all the data which have been considered some indications are present for a time evolution of the dark energy EOS. The joint probability for different couples of parameters which characterize the CPL EOS, are shown in Fig. \ref{likecpl}.
Our statistical analysis has been performed introducing a parametrized redshift variable, the so called \textit{y-redshift}:
\begin{equation}
z \rightarrow y = \frac{z}{1+z}~,
\end{equation}
which maps the z-interval $[0,\infty]$ into the y-interval $[0,1]$ \footnote{This choice facilitates the comparison between the present results and the cosmographic analysis.}.

\begin{figure}
\includegraphics[width=6 cm, height=4 cm]{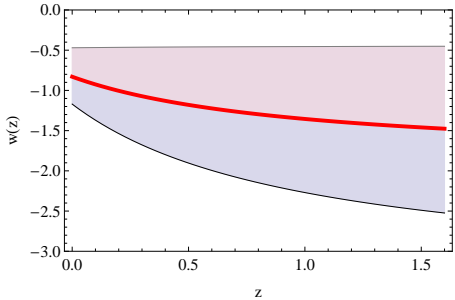}
\caption{ Redshift dependence of the CPL EOS for different values of the EoS parameters $w_0$ and $w_1$. The filled region corresponds to the \textit{allowed} behaviour of
the EOS, when the EOS parameters are varying within the $1 \sigma$ range of confidence. $\Omega_m $ is fixed and
set to the best fit value. The solid red line correspond to $w_0 =w_{0\,bf}$, $w_1 =w_{1\,bf}$.}
\label{wz}
\end{figure}

\begin{table*}
\begin{center}
\resizebox{8cm}{!}{
\begin{tabular}{|cccccc|}
\hline
$Id$ & $x_{bf}$ & $\langle x \rangle$ & $\tilde{x}$ & $68\% \ {\rm CL}$  & $95\% \ {\rm CL}$ \\
\hline
\hline
~ & \multicolumn{5}{c}{$w(z) =w_0 + w_{1} z (1 + z)^{-1} $} \\
\hline
~ & ~ & ~ & ~ & ~ & ~  \\
$\Omega_M$ &0.225 &  0.238 & 0.237 & (0.206, 0.272) & (0.183, 0.305) \\
~ & ~ & ~ & ~ & ~ & ~ \\
$h$ & 0.732 &0.714 & 0.713 &(0.68, 0.745) & (0.659, 0.778)\\
~ & ~ & ~ & ~ & ~ & ~  \\
$w_0$ &-1.15& -0.832 &  -0.834&  (-1.17, -0.476) & (-1.41, -0.36)\\
~ & ~ & ~ & ~ & ~ & ~ \\
$w_1$ & -0.99&-1.06 &-1.05& (-2.2, 0.037) & (-2.8, 0.74)  \\
~ & ~ & ~ & ~ & ~ & ~ \\
\hline
~ & \multicolumn{5}{c}{$w_0+w_1 \left(\frac{\sin (z+1)}{z+1}-\sin (1)\right)$} \\
\hline
~ & ~ & ~ & ~ & ~ & ~  \\
$\Omega_M$ & 0.225 &0.235&0.234 &(0.205, 0.264)& (0.182, 0.294)\\
~ & ~ & ~ & ~ & ~ & ~ \\
$h$ &0.735 & 0.72 & 0.72&(0.69, 0.75)&(0.66, 0.78)\\
~ & ~ & ~ & ~ & ~ & ~ \\
$w_{0}$ & -1.01 &  -0.96& -1.0& (-1.23, -0.742)& (-1.43, -0.493) \\
~ & ~ & ~ & ~ & ~ & ~  \\
$w_{1}$ & 0.14&  0.88 &0.82&(-0.27, 2.1)& (-1.18, 2.8) \\
~ & ~ & ~ & ~ & ~ & ~  \\
\hline
~ & \multicolumn{5}{c}{$ w(z)=\frac{w_1 (1-\cos (\delta  \log (z+1)))}{\log (z+1)}+w_0$} \\
\hline
~ & ~ & ~ & ~ & ~ & ~  \\
$\Omega_M$ & 0.15 &0.154&0.153&(0.15, 0.21)&(0.13, 0.24)\\
~ & ~ & ~ & ~ & ~ & ~ \\
$h$ &0.7 &  0.73 & 0.73&(0.72, 0.75)&(0.7, 0.78)\\
~ & ~ & ~ & ~ & ~ & ~ \\
$w_{0}$ & -1.55 &  -1.54& -1.55&(-1.59, -1.48)& (-1.66, -1.45)\\
~ & ~ & ~ & ~ & ~ & ~  \\
$w_{1}$ & 0.45&0.47 &0.37& (0.07, 0.78)&(0.013, 1.9)\\
~ & ~ & ~ & ~ & ~ & ~  \\
$\delta$ & 0.76&  0.66 &0.62&(0.54, 0.8)& (0.5, 1.1) \\
~ & ~ & ~ & ~ & ~ & ~  \\
\hline
\end{tabular}}
\end{center}
\caption{Constraints on the EOS parameters for different parametrization. Columns report best fit ($x_{bf}$), mean ($\langle x \rangle$) and median ($\tilde{x}$) values and the $68$ and $95\%$ confidence limits.}
\label{tab:fulltab}
\end{table*}
It is well known that the likelihood analysis alone cannot provide an effective way to discriminate between different models. In our analysis we use the so called {\it BIC} as selection criterion\cite{BIC}, defined as
\begin{equation}
BIC=-2 \ln  \mathcal{L}_{max} \, + k \ln N\,,
\end{equation}
where $\mathcal{L}_{max}$, $k$, and $N$ are the maximum likelihood, the number of parameters, which characterizing the models, and the number of data points, respectively. According to this selection criterion, a positive evidence against the model with the higher BIC is defined by a difference $\Delta{\rm BIC}=2$ and a strong evidence is defined by $\Delta{\rm BIC}=6$. Applying such a test to our three parametrization for the EOS, we evaluate $\Delta$BIC for each
model, relative to the CPL model: it turns out that $\Delta BIC>6$, only for the 3D-parametrization $w(z)=\frac{w_1 (1-\cos (\delta  \log (z+1)))}{\log (z+1)}+w_0$, pointing out a strong evidence against this model. In the case of the {\it oscillating  } EOS $w_0+w_1 \left(\frac{\sin (z+1)}{z+1}-\sin (1)\right)$ we find out  $\Delta BIC \simeq 5.9$, underlying a certain (weak) positive evidence against such parametrization.
In Fig. \ref{wz} we show the redshift dependence of the CPL EOS for different values of the EOS parameters $w_0$ and $w_1$, and  in Fig.\ref{likecpl} we show the joint probability for different couple of parameters for the CPL parametrization.
From our investigation we find slight indication for a non-constant EOS, $w$, in any considered parametrization, even if the cosmological constant is not ruled out from these
observations. It turns out that the constraints on the EOS parameters can be strengthen if they are cross-checked with the results of the cosmographic analysis performed on the same datasets in\cite{MECP12}. Without loss of generality, we can, actually,  invert the Eqs. \ref{cosmoeos} for the CPL parametrization, which is favourite by our analysis according to the BIC criterion, and obtain the cosmographic parameters $q_0$ and $j_0$, as functions of the EoS parameters:
\begin{eqnarray}
q_0(\Omega_m, w_0)&=&\frac{1}{2}\left[1+3 w_0(1-\Omega_m)\right]\,,\\ \label{qwo}
j_0(\Omega_m, w_0,w_1)&=& \frac{1}{2} \left[-9 w_0 \left(w_0+1\right) \left(\Omega_m-1\right)+3 w_1
   \left(1-\Omega_m\right)+2\right]\,.\label{jwow1}
\end{eqnarray}

\begin{figure}
\includegraphics[width=6 cm, height=4 cm]{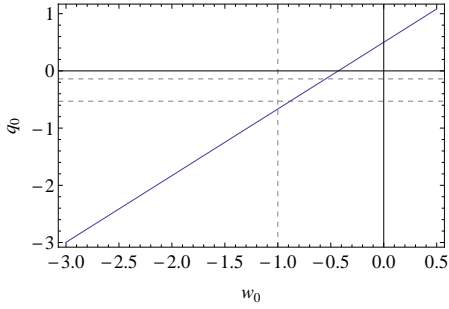}
\caption{Behaviour of $q_0$ as function of $w_0$ for $\Omega_m=0.237$, as results from Eq. (\ref{qwo}).The horizontal dashed lines correspond to the $2 \sigma$ range of confidence for $q_0$.}
\label{qow0cpl}
\end{figure}

\begin{figure}
\includegraphics[width=6 cm, height=4cm]{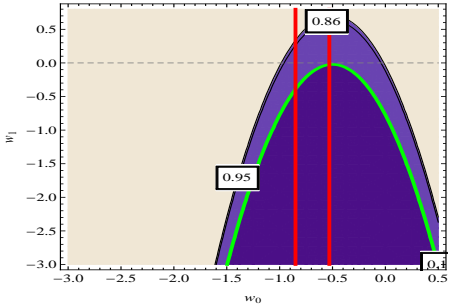}
\caption{Contours plot of the jerk $j_0$ in the plane $w_0, w_1$ for a CPL EOS parametrization, as provided by the cosmographic analysis.
 The value of $\Omega_m$ is set at its median value $\Omega_m=0.237$} \label{jocontoursw0w1}
\end{figure}
\begin{figure}{
       \includegraphics[width=6 cm, height=6 cm]{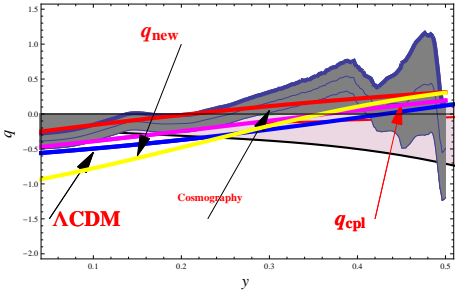}}
        \caption{Reconstruction of deceleration history: the allowed region for $q(z)$,
        obtained by Daly \& Djorgovski, from the full dataset is represented by the shadow area.
        The coloured--solid line  show the deceleration function, $q(z)$ for different EOS parametrization, as indicated by a labe (the label \textit{new} indicates the parametrization in Eq. \ref{noveleos}).
        The red solid line shows the $q(z)$ reconstruction obtained from the cosmography: it is interesting to note that it is all within the region allowed by the data. }
        \label{dalyq}
\end{figure}

In Fig.\ref{qow0cpl} and \ref{jocontoursw0w1} we show the behaviour of  $q_0$ as function of $w_0$ and the contours plot of $j_0$ in the plane $w_0, w_1$ respectively.
If we add  to the Eqs. (\ref{qwo}) and (\ref{jwow1}) the priors on the values of $q_0$ and $j_0$ obtained from the cosmographic analysis, it turns out that the range of confidence of the EOS parameters are {\it squeezed  at } ${\rm 1 \sigma}$ to $w_0 \sim \in \left( -0.87, -0.5 \right)$ and $w_1 \sim \in \left( -0.3, 0.6 \right)$. It is interesting to note, then,  that the use of our  large collection of present data-sets, matched with the cosmographic analysis allows us to improve the constraints on the dark energy EoS competitively with the improvements that can be achieved with future high redshift SNeIa samples (see \cite{salzal2013})   Despite the remarkable improvements in the constraints on the EOS parameters, some caution is needed, due to the circumstance that the systems of algebraic equations in \ref{cosmoeos}, and then its inverse, is highly non linear, and admits multiple solutions for any assigned  n-fold $\left(q_0, j_0,s_0,...\right)$, this resulting in a strong degeneracy among the parameters, which is hard to manage. A possible strategy to ameliorate the maximum likelihood estimates could consists of  incorporating the  restrictions on the EOS parameters, coming from cosmography, in the likelihood itself. For this purpose, in a forthcoming paper we plan to implement at least two possibilities:
\begin{itemize}
\item[i] to use a {\it constrained optimiser} in maximizing the log-likelihood function,
\item[ii]to reparametrized the log-likelihood in such a way that the constraints are eliminated.
\end{itemize}
In \cite{DalyDjorgovski1,DalyDjorgovski2,DalyDjorgovski3} has been developed (and then revised in \cite{Lazkoz:2012eh}), a numerical method for a direct determination, i.e. a determination from the data, of the expansion deceleration parameter, $q(z)$, in terms of the coordinate distance $\mathcal{Y}(z)$, through the equation
\begin{equation}
-q(z) \equiv \ddot{a} a/\dot{a}^2 = 1~ +~ (1+z) ~(d\mathcal{Y}/dz)^{-1}
(d^2 \mathcal{Y}/dz^2)\,,\label{dalyeq}
\end{equation}
valid for flat models. This expression for $q(z)$ is valid for any homogeneous and isotropic Universe in which $(1+z) = a_0/a(t)$, and it is therefore quite general and can be compared with any model to account for the accelerated expansion of the Universe. Using  using the derivation rule
\begin{equation}\label{dty}
\frac{d}{dt} = (1 - y) H \frac{d}{dy}\,.
\end{equation}
we can reconstruct also $q(y)$.
This approach has the advantage to be free from  any assumptions about the nature of the dark energy, but it introduces rather large errors in the estimation of $q(y)$, since the numerical derivation is very sensitive to the size and quality of data. In Fig. (\ref{dalyq}), we compare the $q(y)$ obtained by Daly \& Djorgovski from their full dataset with the $q(z,\mathbf{\theta})$,  reconstructed by using Eq. \ref{q(z)} for any considered EOS parametrization, and also the {\it cosmographic } $q(y)$. It is interesting to note the cosmographically reconstructed $q(y)$ lies completely  within the region allowed by the data.

\section{Discussion and Conclusions}

 In this work  we have presented  constraints on the dark energy EOS obtained  by using an updated collection of observational datasets. In particular we are looking for any indications of a deviation from the $w(z)\neq-1$ come to light,  reflecting  the possibility of a deviation  from the $\Lambda$CDM cosmological model. To accomplish this task we focus on  a direct and full reconstruction of  the dark energy EOS through several parametrizations, widely used in literature.  We have found  indications for a time evolution of the EOS  in any considered parametrization, even if the cosmological constant is not ruled out from these observations. To discriminate between different models, we use the so called {\it BIC} as selection criterion\cite{BIC}: it turns out that the CPL parametrization is favoured by the data. Actually, we  evaluated $\Delta$BIC for each model, relative to the CPL model:  it turns out that $\Delta$$BIC>6$, only for the 3D-parametrization $w(z)=\frac{w_1 (1-\cos (\delta  \log (z+1)))}{\log (z+1)}+w_0$, pointing out a strong evidence against this model. In the case of the {\it oscillating  } EOS $w_0+w_1 \left(\frac{\sin (z+1)}{z+1}-\sin (1)\right)$ we find out  $\Delta BIC \simeq 5.9$, underlying a certain (weak) positive evidence against such parametrization. It turns out tha for the CPL parametrization  $w_0^{median}=-0.834$, and  the
the range of confidence at  ${\rm 1 \sigma}$ is   $w_0 \sim \in \left(-1.17, -0.476 \right)$;   $w_1^{median}=-1.05$ and $w_1 \sim \in \left(-2.2, 0.037 \right)$
Moreover, it turns out that if we include in the  space of parameters $w_0-w_1$ the priors on the values of $q_0$ and $j_0$ obtained from the cosmographic analysis, the constraints on the EOS can be improved competitively with the improvements  achieved with future high redshift SNeIa samples : the  the range of confidence  are, indeed, {\it squeezed  at } ${\rm 1 \sigma}$ to $w_0 \sim \in \left( -0.87, -0.5 \right)$ and $w_1 \sim \in \left( -0.3, 0.6 \right)$. However, since the map connecting the cosmographic and the EOS parameters is highly non linear a strong degeneracy among the parameters is observed, which is hard to manage. Finally,  we reconstruct the  the deceleration parameter $q(z)$ for any considered EOS parametrization, comparing it with the $q(z)$ obtained  from observational dataset, and with  the {\it cosmographic } $q_{cosmographic}(z)$. It is interesting to note that just this $q_{cosmographic}(z)$  lies within the region allowed by the data, thus indicating  that a possible strategy to ameliorate the EOS analysis and  taking into the account the cosmography  results,  could consist of setting up  a sort of {\it constrained maximum likelihood } estimate  within the MCMCs, as we intend to perform in an upcoming paper.

\subsection*{Acknowledgments}

EP, MDL and PS  acknowledge the support of INFN Sez. di Napoli (iniziativa Specifica CQSKY), and also EP and MDL acknowledge the support of INFN Sez. di Napoli  (Iniziativa
Specifica TEONGRAV). MDL is supported by MIUR (PRIN 2009).

\end{document}